\documentclass[12pt]{article} \def\theequation{\arabic{section}.\arabic{equation}}
\newcounter{rown}

\usepackage{mathrsfs,eucal,graphicx}
\usepackage[centertags]{amsmath}
\usepackage{color}
\usepackage{amsfonts}
\usepackage{amssymb}
\usepackage{amsthm}
\usepackage{newlfont}
\theoremstyle{plain}
\newtheorem{lemma}{Lemma}

\newtheorem{theorem}[lemma]{Theorem}

\theoremstyle{definition}

\usepackage[nosort]{cite}
\begin{document}
\renewcommand{\theequation}{\arabic{section}.\arabic{equation}}
\title{\bf On a class of $n$-Leibniz deformations of the
simple Filippov algebras}

\author{ Jos\'e A. de Azc\'{a}rraga, \\
Dept. Theor. Phys. and IFIC (CSIC-UVEG), \\
Univ. of Valencia, 46100-Burjassot (Valencia), Spain\\
 Jos\'e M. Izquierdo,\\
Dept. Theor. Phys., Univ. of Valladolid, \\
47011-Valladolid, Spain}
\date{}
\maketitle
\begin{abstract}
We study the problem of infinitesimal deformations
 of all real, simple, finite-dimensional Filippov (or $n$-Lie)
algebras, considered as a class of $n$-Leibniz algebras characterized by
having an $n$-bracket skewsymmetric in its $n-1$ first arguments. We
prove that all $n>3$ simple finite-dimensional Filippov
algebras are rigid as $n$-Leibniz algebras of this class.
This rigidity also holds for the Leibniz deformations of the
semisimple $n=2$ Filippov ({\it i.e.}, Lie) algebras. The $n=3$ simple
FAs, however, admit a non-trivial one-parameter infinitesimal
3-Leibniz algebra deformation. We also show that the
$n\geq 3$ simple Filippov algebras do not admit non-trivial central
extensions as $n$-Leibniz algebras of the above class.
\end{abstract}
\newpage

\section{Introduction}

Lie algebras can be generalized by relaxing the skewsymmetry of
the Lie bracket. This leads to the {\it Leibniz (or
Loday's) algebras}  $\mathscr{L}$ \cite{Lod:93,Lod-Pir:93,Cuv:94,
Lod-Pir:96}, defined as a vector space $\mathscr{L}$ endowed with a
bilinear operation $\mathscr{L}\times \mathscr{L}\rightarrow
\mathscr{L}$ that satisfies the  {\it Leibniz identity}
\begin{equation}
\label{left-Lei-alg} [X,[Y,Z]]=[[X,Y],Z]+[Y,[[X,Z]] \qquad \forall
X,Y,Z \in \mathscr{L} \quad ,
\end{equation}
which states that $ad_X=[X,\;\,]$ is a derivation of the Leibniz
bracket. Lie algebras $\mathfrak{g}$ are the special class of Leibniz algebras
for which $[X,Y]=-[Y,X]$ $\forall X,Y$. Since the Leibniz algebra
bracket is not skewsymmetric, left and right derivations are not
(anti)equivalent and correspondingly there are two possible versions
of the Leibniz identity; eq.~(\ref{left-Lei-alg}), which we shall
adopt, is the left Leibniz identity and correspondingly defines a
left Leibniz algebra.

Lie algebra deformations \cite{Gers:63,Nij-Rich:67} can be easily
generalized to the Leibniz case. Infinitesimal Leibniz algebra
deformations are defined by a deformed bracket $[X_1,X_2]_t$,
\begin{equation}
\label{def-2-bra}
 [X_1,X_2]_t = [X_1,X_2] + t \alpha (X_1,X_2) \quad ,
 \end{equation}
such that $\alpha(X_1,X_2)$ is a bilinear $\mathscr{L}$-valued map
$\alpha: \mathscr{L} \otimes \mathscr{L} \rightarrow \mathscr{L}\,$,\break
$\,\alpha:(X_1,X_2) \mapsto\alpha(X_1,X_2)$  and $[X_1,X_2]_t$
satisfies (\ref{left-Lei-alg}) (for deformations of Lie algebras
$\mathfrak{g}$, $\alpha$ would be skewsymmetric in its two arguments).
The non-trivial inequivalent infinitesimal deformations of Lie and
Leibniz algebras are classified by the elements of the second
cohomology groups $H^2_{ad}(\mathfrak{g},\mathfrak{g})$ and
$H^2_{ad}(\mathscr{L},\mathscr{L})$ respectively.
 The Leibniz algebra cohomology has been discussed
 in \cite{Lod:93,Lod-Pir:93,Cuv:94,Cas-Lod-Pir:02} (there
for right $\mathscr{L}$s) and in \cite{Da-Tak:97}. The cohomology complex
$(C^\bullet(\mathscr{L},\mathscr{L}),\delta )$ becomes the Lie
algebra one $(C^\bullet(\mathfrak{g},\mathfrak{g}),\delta )$
when $\mathscr{L}= \mathfrak{g}$ and, as a result of the antisymmetry,
the cochains are also required to be antisymmetric. But, since Lie algebras
are also Leibniz, it is also possible to look for Leibniz deformations
of Lie algebras when viewed as Leibniz ones.
This may result in the appearance of more deformations,
a fact recently discussed and observed in \cite{Fia-Man:08}
for the nilpotent 3-dimensional Heisenberg algebra.
In fact, and for a symmetric representation of
$\mathscr{L}$ \cite{Lod:93,Lod-Pir:93}, there is a homomorphism
\cite{Lod-Pir:93} between the Leibniz and Lie algebra
homologies as well as between the Lie algebra and Leibniz
cohomologies for that representation.

Similarly, one may consider central extensions of Leibniz
algebras. Given a real Leibniz algebra $\mathscr{L}$ with a
basis $\{X_a\}$, a central extension $\widetilde{ \mathscr{L}}$ is
given by the vector space spanned by the vectors $\widetilde{X}_a$
plus an additional central generator $\Xi$, endowed with the bracket
\begin{equation}
\label{intro-1}
  [\widetilde{X}_a,\widetilde{X}_b] = C^c_{ab} \widetilde{X}_c  +
  \omega(X_a,X_b) \Xi \ ,
  \quad [\widetilde{X}_a,\Xi]
  = 0 = [\Xi,\widetilde{X}_a]
\end{equation}
where the structure constants at the $l.h.s.$ are those of the
unextended $\mathscr{L}$ and $\omega(X_a,X_b)$ is an
$\mathbb{R}$-valued bilinear map which, in contrast with the Lie
algebra case, does not have to be antisymmetric in its arguments.
This also follows from the fact that a central extension of a
Leibniz algebra may be viewed as  a deformation of the direct sum
of $\mathscr{L}$ and the one-dimensional algebra generated by
$\Xi$ that keeps the central character of $\Xi$.

Other generalizations of the Lie algebra structure follow by considering
brackets with $n>2$ entries. A first $n$-ary generalization is
obtained by extending the derivation property reflected
by the ordinary ($n=2$) Jacobi identity (JI) to the $n$-bracket.
This leads to the $n$-{\it Lie} or {\it Filippov}
algebras $\mathfrak{G}$ \cite{Filippov},\cite{Kas:87, Ling:93}
(other $n$-ary generalizations, based on the fact that the
JI also follows from associativity, are possible \cite{JPA,CMP,HW95}).
Filippov or $n$-Lie algebras (FAs) are given by a vector
space $\mathfrak{G}$ endowed with a skew-symmetric, $n$-linear bracket,
\begin{equation}
       (X_1,\dots , X_n)\in \mathfrak{G} \times\dots \times  \mathfrak{G}
       \mapsto [X_1,\dots , X_n] \in \mathfrak{G}
\label{intro1}
\end{equation}
that satisfies the {\it Filippov identity} (FI),
\begin{equation}
\label{intro2}
       [X_1,\dots , X_{n-1},[Y_1,\dots Y_n]] = \sum_{a=1}^n
       [Y_1,\dots Y_{a-1}, [X_1,\dots , X_{n-1}, Y_a ], Y_{a+1}
       ,\dots Y_n] \;
\end{equation}
which, for $n=2$ is the JI; clearly, a 2-Lie algebra is an ordinary Lie one.
The FI  states that the left action of the linear operator
$[X_1,\dots,X_{n-1},\;\,]\equiv ad_{\mathscr{X}} \in \textrm{End}\; \mathfrak{G}$,
defined by
\begin{equation}
\label{intro2a}
ad_{\mathscr{X}} Z:= [X_1,\dots , X_{n-1},Z]\equiv \mathscr{X}\cdot Z
\qquad \forall Z\in \mathfrak{G} \; ,
\end{equation}
where $\mathscr{X}=(X_1,\dots,X_{n-1})
\in \wedge^{n-1} \mathfrak{G}$ , is a derivation of the FA.
It is convenient to have a name for the elements
$\mathscr{X}\in \wedge^{n-1} \mathfrak{G}$ that define inner derivations
$\mathscr{X}\cdot \equiv ad_\mathscr{X}$ of $\mathfrak{G}$,
since they will reappear when defining
the different cohomology complexes; they are called
\cite{FAcoho, review} `fundamental objects' of the Filippov
algebra. Detailed accounts of the properties of FAs are given in
\cite{Filippov,Kas:87,Ling:93} (see \cite{review} for a review).
FAs have been considered in physics in the
context of Nambu mechanics \cite{Nambu:73,Tak:93} and, recently
(for $n=3$), in the search for the effective action of coincident
M2-branes in M-theory initiated by the Bagger-Lambert-Gustavsson
(BLG) model \cite{Ba-La:07b,Gustav:08} (further
references on the physical applications of $n$-ary algebras
are given in \cite{review}).

The skewsymmetry of the FA $n$-bracket may also be relaxed. This
gives rise to the $n$-{\it Leibniz algebras} $\mathfrak{L}$
\cite{Da-Tak:97,Cas-Lod-Pir:02} (for $n=2$, Leibniz
algebras $\mathscr{L}$). The left $n$-Leibniz algebras are
defined by eqs.~(\ref{intro1}), (\ref{intro2}) without requiring
skewsymmetry for the $n$-bracket. As a result, the fundamental
objects of $\mathfrak{L}$ are no longer skewsymmetric and
$\mathscr{X}\in\otimes^{n-1}\mathfrak{L}$ in general. As for FAs,
one may consider {\it infinitesimal deformations of $n$-Leibniz
algebras} $\mathfrak{L}$. Then one is led to the study of the
$n$-Leibniz algebra cohomology for the adjoint action $ad$. The
expression that gives the action of the coboundary operator
(Sec.~\ref{def-n-L-co}) is the same as the one suitable
to study FA deformations, the only difference
being that antisymmetry is not required in the $n$-Leibniz case
(in fact, since both FA and $n$-Leibniz cohomologies
are based on the FI \eqref{intro2},
$n$-Leibniz cohomology underlies FA cohomology). We
will see in Sec.~\ref{def-n-L-co} that the non-equivalent infinitesimal
deformations of a given $n$-Leibniz algebra are in one-to-one
correspondence with the elements of the first cohomology group
$H^1_{ad}(\mathfrak{L},\mathfrak{L})$.

Besides deformations, one may
consider {\it central extensions of an $n$-Leibniz algebra} by
generalizing eq.~\eqref{intro-1} for the Leibniz $n$-bracket. The
relevant cohomology here is formally the same as
for the central extensions of FAs: central extensions of $n$-Leibniz
algebras are classified by the first cohomology group
$H^1(\mathfrak{L},\mathbb{R})$, as  will be seen in Sec.~\ref{n-L-central}.
As before for deformations, the fact that it is
the first cohomology group that matters (rather than the second,
as it would be for extensions of Leibniz algebras $\mathscr{L}$
\cite{Lod:93,Lod-Pir:93,Cuv:94} with the standard counting
for $n=2$ algebras), is just a notational consequence of the
natural labelling of $p$-cochains for general $n$-ary
algebras, a point made clear in Sec.~\ref{def-n-L-co} where
their deformation cohomology is presented.

Since all FAs are, in particular, $n$-Leibniz algebras with fully
anticommuting $n$-brackets, one may again consider
$n$-{\it Leibniz} infinitesimal deformations and central extensions
of $n$-Lie algebras viewed as $n$-Leibniz ones.
A FA may admit a non-trivial $n$-Leibniz
deformation when looked at as an $n$-Leibniz algebra, even if the
original FA is rigid under FA deformations. For instance, it has
been proven \cite{FAcoho} that a Whitehead Lemma holds for all
semisimple $n$-Lie algebras. As a result, all simple (in fact
semisimple) FAs are rigid under FA deformations \cite{FAcoho}.
However, the simple $n=3$ Euclidean FA $A_4$
admits a non-trivial 3-Leibniz deformation \cite{F-O'F:08}. A
natural question to ask, which we shall address in this paper, is
whether there exist non-trivial $n$-Leibniz deformations
of the finite-dimensional simple FAs. Similarly, it is also natural
to look for non-trivial $n$-Leibniz central extensions of simple
FAs, in spite of the fact that the all semisimple FA central
extensions are known to be trivial by the above extension to all $n$-Lie
algebras \cite{FAcoho} of the well known $n=2$ Whitehead Lemma.

We shall study the above two problems, infinitesimal deformations
and central extensions, for FAs considered as a particular
case of $n$-Leibniz algebras, the class that keeps the antisymmetry
of the first $n-1$ entries in the $n$-bracket and thus
has fundamental objects $\mathscr{X}$ that
are skewsymmetric (see eq.~\eqref{intro2a}). For
$n=3$, examples of this type of real Leibniz
algebras have appeared in the study of multiple M2-branes
\cite{Cher-Sa:08,Cher-Do-Sa:08} (other examples weakening
the skewsymmetry have been considered in physics, such as the `hermitean'
algebras \cite{Bag-Lam:08} which will not be considered here;
see also \cite{Gra-Nil-Pe:08}). These non-fully
commutative 3-algebras correspond to 3-Leibniz algebras
with a 3-bracket that retains the skewsymmetry for the first
two arguments or, equivalently, to 3-Leibniz algebras for which the
fundamental objects $\mathscr{X}=(X_1,X_2)$ are still antisymmetric as in the
FA case; the 3-Leibniz infinitesimal deformation of the FA $A_4$
in \cite{F-O'F:08} is of this class. We shall show that this
is, in fact, the only 3-Leibniz deformation and, further, that there
are no non-trivial deformations of the above type when $n\neq 3$.
The semisimple $n=2$ FAs $\mathfrak{G}=\mathfrak{g}$, for which the
fundamental objects are single elements of the Lie algebra
$\mathfrak{g}$ (hence with no restrictions), will also turn out
to be Leibniz rigid. Finally, we will also prove
for simple $n\geq 3$ FAs that there are no non-trivial
central $n$-Leibniz extensions of the mentioned class.

The plan of the paper is the following: in Sec.~\ref{def-ext} we look at the
deformation and the central extension theory of $n$-Leibniz
algebras, and give the one-cocycle and one-coboundary conditions for
the appropriate cohomologies both in their intrinsic forms and in
coordinates. Sec.~\ref{n-L-defSec} proves that there are no $n$-Leibniz algebra
deformations with $n$-brackets skewsymmetric in the first $n-1$
entries (and thus with fully skewsymmetric fundamental objects) of
simple $n$-Lie algebras when $n>3$. This result will be obtained
in Sec.~\ref{sec-dual} using the coordinate expressions given in Sec.~\ref{inf-def-sub};
the proof for the $n=2$ case is given in Sec.~\ref{Leib-def-of-Lie}.
Sec.~\ref{Leib-central-Lie} follows a similar procedure to prove
that $n \geq 3$ simple FAs do not have non-trivial $n$-Leibniz central
extensions of the mentioned class.

  All algebras in this paper are real and finite-dimensional. Our results
also hold in the complex case, but we prefer to consider real algebras
since moving to $\mathbb{C}$ does not allow us to distinguish between
the different physically interesting pseudoEuclidean simple FAs
(see eq.~\eqref{forleib8}). Theorems 1 (Sec.~\ref{Leib-def-of-Lie}) and 2
(Sec.\ref{Leib-central-Lie}) are our results; Sec.~\ref{final-remarks}
comments on possible extensions of the present work.

\section{Deformations and extensions of $n$-Leibniz algebras and cohomology}
\label{def-ext}

\subsection{Infinitesimal deformations  of $n$-Leibniz algebras}
\label{inf-def-sub}
Let $\mathfrak{L}$ be an $n$-Leibniz algebra. Its $n$-bracket obeys
the $n$-Leibniz identity which, in fact, is formally identical to
the FI (\ref{intro2}), the only difference being that now the
$n$-Leibniz bracket need not be fully skewsymmetric. A one-parameter
infinitesimal deformation of $\mathfrak{L}$ is given by a new,
deformed, bracket
\begin{equation}
\label{ntacohomology1}
    [X_1 ,\dots , X_n]_t = [X_1 ,\dots , X_n] +t \alpha^1(X_1 ,\dots ,
    X_n) \ ,
\end{equation}
where $\alpha^1$ is a linear $\mathfrak{L}$-valued map
$\alpha^1: \mathfrak{L} \otimes \mathop{\cdots}\limits^{n-1} \otimes \mathfrak{L} \otimes
\mathfrak{L} \rightarrow \mathfrak{L}$ (an $\mathfrak{L}$-valued
{\it one}-cochain, as we shall see at the end of the
section) and $t$ is the parameter of
the infinitesimal deformation.

  The requirement that the deformed bracket also obeys
the $n$-Leibniz identity (the FI) leads to a condition
on $\alpha^1$ that may be interpreted as
the one-cocycle condition in the cohomology for the deformation of
$n$-Leibniz algebras. Namely, it is $\delta \alpha^1 =0$ with
\begin{eqnarray}
\label{forleib1}
 & & \delta \alpha^1(X_1, \dots, X_{n-1},Y_1, \dots ,Y_{n-1}, Z)
 \nonumber \\
  & & \quad \quad = [X_1, \dots, X_{n-1},
\alpha^1(Y_1, \dots ,Y_{n-1},Z)] +
\alpha^1(X_1, \dots, X_{n-1}, [Y_1, \dots ,Y_{n-1},Z]) \nonumber\\
 & & \quad \quad - \sum^{n-1}_{r=1} [Y_1, \dots ,Y_{r-1},
\alpha^1(X_1, \dots, X_{n-1},Y_r),Y_{r+1} ,\dots , Y_{n-1},Z]
\nonumber\\
& & \quad \quad -[Y_1, \dots ,Y_{n-1},\alpha^1(X_1, \dots,
X_{n-1},Z)]
\nonumber\\
& & \quad \quad - \sum^{n-1}_{r=1} \alpha^1(Y_1, \dots ,Y_{r-1},
[X_1, \dots, X_{n-1},Y_r],Y_{r+1} ,\dots , Y_{n-1},Z) \nonumber\\
& & \quad\quad - \alpha^1(Y_1, \dots ,Y_{n-1}, [X_1, \dots,
X_{n-1},Z]) \; .
\end{eqnarray}
The $\delta \alpha^1 =0$ condition above is the same as for FA deformations,
but now there are no antisymmetry conditions for the cochains.  In
terms of the {\it fundamental objects} of the $n$-Leibniz algebra
$\mathfrak{L}$, where now $\mathscr{X}=(X_1,\dots X_{n-1})\in \otimes^{n-1}\mathfrak{L}$,
eq.~\eqref{forleib1} may be rewritten as \cite{FAcoho,review}
\begin{equation}
\label{def-1-coch-b}
\begin{aligned}
(\delta\alpha^1)(\mathscr{X}, \mathscr{Y}, Z) = ad_{\mathscr{X}}&
\alpha^1(\mathscr{Y}, Z) -ad_{\mathscr{Y}}\alpha^1(\mathscr{X}, Z)
  -(\alpha^1(\mathscr{X}, \quad )\cdot \mathscr{Y})\cdot Z \\
 -\alpha^1 (\mathscr{X} & \cdot\mathscr{Y}, Z)  -\alpha^1(\mathscr{Y}, \mathscr{X}\cdot Z)
 + \alpha^1(\mathscr{X}, \mathscr{Y}\cdot Z) = 0 \quad ,
 \end{aligned}
\end{equation}
where, for instance for $n=3$, the term
$\alpha^1(\mathscr{X},\quad)\cdot \mathscr{Y}$ above is the
fundamental object defined by
\begin{equation}
\label{dosf}
\begin{aligned}
 \alpha^1(\mathscr{X},\quad)\cdot \mathscr{Y} :=&
 (\alpha^1(\mathscr{X},\quad)\cdot Y_1,\,Y_2) \,+\, (Y_1,\,\alpha^1(\mathscr{X},\quad)\cdot Y_2)\\
 =& (\alpha^1(\mathscr{X},Y_1), Y_2) \,+\, (Y_1,\alpha^1(\mathscr{X},Y_2)) \quad
 ,\\
&[\,\alpha^1(\mathscr{X},\quad)\cdot Y_i
:=\alpha^1(\mathscr{X},Y_i)\,] \quad ,
\end{aligned}
\end{equation}
$\mathscr{X}\cdot Z = [\mathscr{X},Z] \equiv [X_1,\dots , X_{n-1},Z]$ and
\begin{eqnarray}
\label{intro4}
        \mathscr{X}\cdot \mathscr{Y} := \sum_{a=1}^{n-1}
        (Y_1,\dots ,Y_{a-1} ,[X_1,\dots , X_{n-1}, Y_a], Y_{a+1},
        \dots, Y_{n-1})
\end{eqnarray}
above defines the composition of fundamental objects.

Let us choose a basis $\{ X_a \}$ of $\mathfrak{L}$ for which
\begin{eqnarray}\label{forleib2}
   [X_{a_1},\dots ,X_{a_n}] &=& {f_{a_1 \dots a_n}}^b X_b \quad, \\
\alpha^1(X_{a_1},\dots ,X_{a_n}) &=& X_b {(\alpha^1)^b}_{a_1 \dots
a_n} \quad .
\end{eqnarray}
Then, the one-cocycle condition \eqref{forleib1} for the
one-cochain $\alpha^1$ takes the form
\begin{eqnarray}
\label{forleib3}
 & & {f_{a_1 \dots a_{n-1}e}}^d {(\alpha^1)^e}_{b_1 \dots b_{n-1}c}
+ {(\alpha^1)^d}_{a_1 \dots a_{n-1}e} {f_{b_1 \dots b_{n-1}c}}^e
\nonumber\\
& & -  \sum^{n-1}_{r=1} {f_{b_1 \dots b_{r-1}eb_{r+1}\dots b_{n-1}
c}}^d
 {(\alpha^1)^e}_{a_1 \dots a_{n-1}b_r} \nonumber\\
& &- {f_{b_1 \dots b_{n-1}e}}^d {(\alpha^1)^e}_{a_1 \dots
a_{n-1}c}
\nonumber\\
& & -  \sum^{n-1}_{r=1} {(\alpha^1)^d}_{b_1 \dots
b_{r-1}eb_{r+1}\dots b_{n-1}c}
{f_{a_1 \dots a_{n-1}b_r}}^e \nonumber\\
& & - {(\alpha^1)^d}_{b_1 \dots b_{n-1}e} {f_{a_1 \dots
a_{n-1}c}}^e
 = 0 \; .
\end{eqnarray}
When $n=3$, this reduces to
\begin{eqnarray}\label{forleib4}
& &  {f_{a_1a_2e}}^d {(\alpha^1)^e}_{b_1b_2c} +
{(\alpha^1)^d}_{a_1a_2e}
{f_{b_1b_2c}}^e \nonumber\\
& & - {f_{eb_2c}}^d {(\alpha^1)^e}_{a_1a_2b_1}- {f_{b_1ec}}^d
{(\alpha^1)^e}_{a_1a_2b_2} \nonumber\\
& & -{f_{b_1b_2e}}^d {(\alpha^1)^e}_{a_1a_2c} \nonumber\\
& &-{(\alpha^1)^d}_{eb_2c} {f_{a_1a_2b_1}}^e -
{(\alpha^1)^d}_{b_1ec}
{f_{a_1a_2b_2}}^e \nonumber\\
& &- {(\alpha^1)^d}_{b_1b_2e} {f_{a_1a_2c}}^e = 0 \ .
\end{eqnarray}

  An $n$-Leibniz algebra deformation is trivial if there is a
redefinition of the generators $X'_i=X_i-t\alpha^0(X_i)$, defined
by some $\mathfrak{L}$-valued zero-cochain $\alpha^0$,
$\alpha^0: \mathfrak{L} \rightarrow \mathfrak{L}$, that removes
the deforming term in eq.~\eqref{ntacohomology1}. This means that
$\alpha^1$ is actually a one-coboundary {\it i.e.}
$\alpha^1=\delta \alpha_0$ since then
\begin{eqnarray}
\label{forleib5}
  & & \alpha^1(X_1,\dots ,X_{n-1},Z) =
 -\alpha^0([X_1,\dots ,X_{n-1},Z]) \nonumber\\
 \quad &+& \sum^{n-1}_{r=1}
[X_1,\dots ,X_{r-1},\alpha^0(X_r),X_{r+1},\dots ,X_{n-1},Z]
 + [X_1,\dots ,X_{n-1},\alpha^0(Z)]  \nonumber\\
\quad &=&\delta \alpha^0(X_1,\dots ,X_{n-1},Z) \; .
\end{eqnarray}
Again, this condition is the same
(but for the skewsymmetry  of both the bracket and the cochain in its arguments)
as the one that establishes that an infinitesimal FA deformation is trivial.
In terms of fundamental objects it is written as
\begin{equation}
\label{gval-one-cob}
 (\alpha^1)(\mathscr{X},Z)\equiv (\delta \alpha^0) (\mathscr{X},Z)
 = \mathscr{X}\cdot \alpha^0(Z)
 -\alpha^0(\mathscr{X}\cdot Z)
 + (\alpha^0(\quad)\cdot \mathscr{X})\cdot Z \quad .
\end{equation}
In the basis $\{ X_a \}$, the coordinates ${(\alpha^0)^b}_a$ of
$\alpha^0$ are defined through
\begin{equation}
\label{forleib6}
\alpha^0(X_a)= {(\alpha^0)^b}_a X_b \; ,
\end{equation}
and the one-coboundary condition \eqref{gval-one-cob} that
expresses $\alpha^1$ in terms of $\alpha^0$ is given by
\begin{eqnarray}
\label{forleib6c}
 {(\alpha^1)^b}_{a_1\dots a_{n-1}c} & = & - {(\alpha^0)^b}_s
{f_{a_1\dots a_{n-1}c}}^s \nonumber\\
 & & + \sum^{n-1}_{r=1} {f_{a_1\dots a_{r-1}sa_{r+1} \dots a_{n-1}c}}^b
{(\alpha^0)^s}_{a_r} \nonumber\\
& & + {f_{a_1\dots a_{n-1}s}}^b {(\alpha^0)^s}_c \ .
\end{eqnarray}
For a 3-Leibniz algebra this gives
\begin{eqnarray}
\label{forleib7}
{(\alpha^1)^b}_{a_1a_2c} &=& -{(\alpha^0)^b}_s {f_{a_1a_2c}}^s
\nonumber\\
& & + {f_{sa_2c}}^b {(\alpha^0)^s}_{a_1} + {f_{a_1sc}}^b
{(\alpha^0)^s}_{a_2}
\nonumber\\
& & {f_{a_1a_2s}}^b {(\alpha^0)^s}_c \ .
\end{eqnarray}

\subsection{$n$-Leibniz algebra deformations cohomology complex}
\label{def-n-L-co}

The action of the coboundary operator $\delta$  on  zero- and
one-cochains given above can be extended to arbitrary
$p$-cochains $\alpha^p$ so that, in Gerstenhaber's sense,
the deformation theory of $n$-Leibniz algebras generates
the corresponding cohomology. In it, $p$-cochains are defined as
elements $\alpha^p \in \textrm{Hom}(\otimes^{p(n-1)+1}
\mathfrak{L}, \mathfrak{L})$; thus, a $p$-cochain takes
$p(n-1)$+$1$ arguments in $\mathfrak{L}\,$, $\,p(n-1)$ of which
enter through $p$ fundamental objects and $\delta\alpha^p$ has
order ($p$+1). The previous expressions
for $\delta\alpha^0$, $\delta\alpha^1$ now generalize to
provide the action of the $n$-Leibniz deformations cohomology coboundary
operator $\delta$ on an arbitrary $p$-cochain $\alpha^p$. This
action is best expressed in terms of fundamental objects,
which explains why $p$ and $p$+$1$ determine their
order (see \cite{FAcoho,review}).

  This leads to the $n$-{\it Leibniz algebra deformation cohomology \break
complex}
$(C^\bullet_{ad}(\mathfrak{L},\mathfrak{L}),\delta)$, where
\begin{equation}
\begin{aligned}
\label{adcoho} (\delta\alpha^p) &
(\mathscr{X}_1,\dots,\mathscr{X}_p,\mathscr{X}_{p+1},Z)= \\
&\sum_{1\leq j<k}^{p+1} (-1)^j
\alpha^p(\mathscr{X}_1,\dots,\widehat{\mathscr{X}_j},\dots,\mathscr{X}_{k-1},
\mathscr{X}_j\cdot
\mathscr{X}_k,\mathscr{X}_{k+1},\dots,\mathscr{X}_{p+1}, Z)\\
+& \sum_{j=1}^{p+1} (-1)^j \alpha^p
(\mathscr{X}_1,\dots,\widehat{\mathscr{X}_j},\dots,\mathscr{X}_{p+1},\mathscr{X}_j\cdot
 Z) \\
 + &\sum_{j=1}^{p+1} (-1)^{j+1} \mathscr{X}_j \cdot
 \alpha^p(\mathscr{X}_1,\dots,\widehat{\mathscr{X}_j}\dots,\mathscr{X}_{p+1},
 Z) \\
 +&(-1)^{p}
 (\alpha^p(\mathscr{X}_1,\dots,\mathscr{X}_p\,,\quad)\cdot
 \mathscr{X}_{p+1})\cdot Z
\end{aligned}
\end{equation}
where, in the last term (see \cite{FAcoho,review}),
\begin{equation}
\label{p-coc-ftal}
 \alpha^p(\mathscr{X}_1,\dots,\mathscr{X}_p\,,\quad)\cdot
 \mathscr{Y}=
 \sum_{i=1}^{n-1}
 (Y_1,\dots,\alpha^p(\mathscr{X}_1,\dots,\mathscr{X}_p,Y_i),\dots,Y_{n-1})
 \; ;
\end{equation}
note that both the left and the right actions
intervene. The nilpotency $\delta^2=0$ is guaranteed by the
FI \eqref{intro2} for $n$-Leibniz algebras.\\

Consequently, we see that the infinitesimal deformations of
$n$-Leibniz algebras are governed  by the non-trivial elements of the
first cohomology group $H^1_{ad}(\mathfrak{L},\mathfrak{L})$
\footnote{When $n=2$, $\mathfrak{L}=\mathscr{L}$
and a one-cochain contains two Leibniz algebra $\mathscr{L}$
arguments. Thus, if its order were determined by this number (as it
is usually the case when $n=2$), it would be a two- rather than a one-cochain.}.

The above cohomology complex is essentially equivalent to that
previously given by Gautheron \cite{Gau:96} in the context of
Nambu algebras; see further \cite{Da-Tak:97,Tak:95,Rot:05}.

\subsection{Central extensions of $n$-Leibniz algebras}
\label{n-L-central}

The central extensions of an $n$-Leibniz algebra $\mathfrak{L}$ are
obtained by adding to the $\mathfrak{L}$ generators a central one.
Thus, the $n$-brackets of the extended algebra are
\begin{eqnarray}
\label{central-ext}
  [\widetilde{X}_1,\dots ,\widetilde{X}_n] &=& {f_{a_1\dots a_n}}^b
  \widetilde{X}_b + \alpha^1(X_1,\dots
  ,X_n)\Xi \ ,\nonumber\\
  \left[\widetilde{X}_1,\dots, \widetilde{X}_{i-1} ,\Xi , \widetilde{X}_{i+1},\dots,
  \widetilde{X}_n\right] &=& 0 \; , \; i=1,\dots n \quad .
\end{eqnarray}
 In contrast with the FA case, the $\mathbb{R}$-valued $n$-Leibniz one-cochain
$\alpha^1$ does not have to be skewsymmetric in its arguments.

As before, a one-cocycle condition arises
for $\alpha^1$ when imposing that the centrally extended algebra is an
$n$-Leibniz one {\it i.e.}, that it obeys the (left) Filippov identity.
In terms of the fundamental objects $\delta\alpha^1=0$ reads
\begin{equation}
\label{1coc-ext-f} (\delta\alpha^1)(\mathscr{X}, \mathscr{Y}, Z) =
 -\alpha^1 (\mathscr{X}  \cdot\mathscr{Y}, Z)  -\alpha^1(\mathscr{Y}, \mathscr{X}\cdot Z)
 + \alpha^1(\mathscr{X}, \mathscr{Y}\cdot Z) = 0 \; ,
\end{equation}
Using eq. \eqref{forleib2}, the coordinates expression
for the one-cocycle condition is given by
\begin{eqnarray}
\label{1coc-coord}
 {(\alpha^1)}_{a_1 \dots a_{n-1}e} {f_{b_1 \dots b_{n-1}c}}^e
&-&  \sum^{n-1}_{r=1} {(\alpha^1)}_{b_1 \dots b_{r-1}eb_{r+1}\dots b_{n-1}c}
{f_{a_1 \dots a_{n-1}b_r}}^e \nonumber\\
 & -& {(\alpha^1)}_{b_1 \dots b_{n-1}e} {f_{a_1 \dots
a_{n-1}c}}^e
 = 0 \ .
\end{eqnarray}

An $n$-Leibniz algebra central extension is trivial if there is a
redefinition of its generators
$\widetilde{X}'_i= \widetilde{X}_i-t\alpha^0(X_i)$
that removes the central term in eq.~\eqref{central-ext}.
In this case $\alpha^1$ satisfies
\begin{equation}
\label{1cob-ext}
   \alpha^1(X_1,\dots ,X_{n-1},Z) = \delta \alpha^0(X_1,\dots
  ,X_{n-1},Z)=  -\alpha^0([X_1,\dots ,X_{n-1},Z]) \ ,
\end{equation}
where $\alpha^0$ is the zero-cochain that generates the
one-coboundary $\alpha^1$. Again, this condition is
formally the same that establishes that a
FA central extension is trivial, although now
$\alpha^1 \in \mathrm{Hom} (\otimes^{n-1}
\mathfrak{L} \otimes \mathfrak{L}, \mathbb{R})$ (rather than
$\alpha^1 \in \mathrm{Hom} (\wedge^{n-1} \mathfrak{G} \wedge
\mathfrak{G}, \mathbb{R})$ as for a FA $\mathfrak{G}$).
In terms of fundamental objects eq.~\eqref{1cob-ext} reads
simply
\begin{equation}
\label{1cob-ext-f}
 (\alpha^1)(\mathscr{X},Z)=
 -\alpha^0(\mathscr{X}\cdot Z)
 \quad .
\end{equation}
Using the basis (\ref{forleib2}) and
\begin{equation}
\label{alpha-0-ext}
\alpha^0(X_a)= \alpha^0{}_a  \ ,
\end{equation}
the coboundary condition $\alpha^1=\delta \alpha^0$ relates
the coordinates of $\alpha^1$ to those of $\alpha^0$ by
\begin{equation}
\label{cob-ext-co}
 (\alpha^1)_{a_1\dots a_{n-1}c} =  - {(\alpha^0)}_s
{f_{a_1\dots a_{n-1}c}}^s  \ .
\end{equation}

As before, the action of the coboundary operator $\delta$  given
above on the  zero- and one-cochains of the central extension
problem can be extended to arbitrary cochains
$\alpha^p\in\hbox{Hom}(\otimes^{p(n-1)+1}\mathfrak{L},
\mathfrak{L})$; the nilpotency of $\delta$ is satisfied by virtue of
the FI \eqref{intro2} for $\mathfrak{L}$. This leads to (see
\cite{FAcoho,review})
\begin{equation}
\begin{aligned}
\label{centralcoho} (\delta\alpha^p) &
(\mathscr{X}_1,\dots,\mathscr{X}_p,\mathscr{X}_{p+1},Z)= \\
&\sum_{1\leq j<k}^{p+1} (-1)^j
\alpha^p(\mathscr{X}_1,\dots,\widehat{\mathscr{X}_j},\dots,\mathscr{X}_{k-1},
\mathscr{X}_j\cdot
\mathscr{X}_k,\mathscr{X}_{k+1},\dots,\mathscr{X}_{p+1}, Z)\\
+& \sum_{j=1}^{p+1} (-1)^j \alpha^p
(\mathscr{X}_1,\dots,\widehat{\mathscr{X}_j},\dots,\mathscr{X}_{p+1},\mathscr{X}_j\cdot
 Z) \quad ,
\end{aligned}
\end{equation}
which determines the cohomology complex
$(C_0^\bullet(\mathfrak{L},\mathbb{R}),\delta)$. It follows that
the central extensions of an $n$-Leibniz algebra are governed by
the first cohomology group $H^1_0(\mathfrak{L},\mathbb{R})$.

  Note that all the expressions corresponding to the central
extensions case can be obtained from those of the deformation
cohomology by using $\mathbb{R}$-valued cochains in place of
$\mathfrak{L}$-valued ones and by substituting the trivial
action for the adjoint one. In coordinates this means removing the
first, upper index of ${(\alpha^1)^d}_{a_1\dots a_{n-1}b}$ and
ignoring all terms containing the $ad$ action in the expression of
$\delta$ in eqs. \eqref{def-1-coch-b}, \eqref{forleib3}.

\section{A class of $n$-Leibniz algebra
deformations of the real simple Filippov algebras}
\label{n-L-defSec}

\subsection{The $n\geq 3$ simple FAs case}
\label{Leib-def-FAn>2}

Consider now the  simple real finite-dimensional Filippov algebras.
These were given in \cite{Filippov} and found to be the only ones in
\cite{Ling:93}. They are constructed on $(n+1)$-dimensional vector
spaces and are characterized by the structure constants
\cite{Filippov, Ling:93}
\begin{equation}
\label{forleib8}
{f_{a_1\dots a_{n-1}c}}^b = (-1)^n \varepsilon_b {\epsilon_{a_1\dots
a_{n-1}c}}^b \quad , \quad a,b,c=1,\dots (n+1) \; ,
\end{equation}
where $\epsilon$ is the Euclidean ($n+1$)-dimensional skewsymmetric
tensor and, following Filippov's notation,  the
$\varepsilon_b$ (no sum in $b$) are just signs that appear in the
pseudoeuclidean algebras and that are absent when considering
the Euclidean $(n+1)$-dimensional simple $n$-Lie
algebra $A_{n+1}$.

Thus, the one-cocycle  condition relevant for the deformation
of a simple $n$-Lie algebra follows from eq.~\eqref{forleib3}
and is given by
\begin{eqnarray}
\label{forleib9}
  & & \varepsilon_d {\epsilon_{a_1 \dots a_{n-1}e}}^d {(\alpha^1)^e}_{b_1 \dots b_{n-1}c}
+\varepsilon_e {(\alpha^1)^d}_{a_1 \dots a_{n-1}e} {\epsilon_{b_1
\dots b_{n-1}c}}^e
\nonumber\\
& & -  \sum^{n-1}_{r=1}  \varepsilon_d {\epsilon_{b_1 \dots
b_{r-1}eb_{r+1}\dots b_{n-1} c}}^d
 {(\alpha^1)^e}_{a_1 \dots a_{n-1}b_r} \nonumber\\
& &- \varepsilon_d {\epsilon_{b_1 \dots b_{n-1}e}}^d
{(\alpha^1)^e}_{a_1 \dots a_{n-1}c}
\nonumber\\
& & -  \sum^{n-1}_{r=1}\varepsilon_e {(\alpha^1)^d}_{b_1 \dots
b_{r-1}eb_{r+1}\dots b_{n-1}c}
{\epsilon_{a_1 \dots a_{n-1}b_r}}^e \nonumber\\
& & - \varepsilon_e {(\alpha^1)^d}_{b_1 \dots b_{n-1}e}
{\epsilon_{a_1 \dots a_{n-1}c}}^e
 = 0  \quad .
\end{eqnarray}
A one-cocycle for a simple $n$-Lie algebra is
actually a one-coboundary (eq.~\eqref{forleib6c}) when
\begin{eqnarray}\label{forleib10}
 {(\alpha^1)^b}_{a_1\dots a_{n-1}c} & = & -\varepsilon_s {(\alpha^0)^b}_s
{\epsilon_{a_1\dots a_{n-1}c}}^s \nonumber\\
 & & + \sum^{n-1}_{r=1} \varepsilon_b
{\epsilon_{a_1\dots a_{r-1}sa_{r+1} \dots a_{n-1}c}}^b
{(\alpha^0)^s}_{a_r} \nonumber\\
& & + \varepsilon_b {\epsilon_{a_1\dots a_{n-1}s}}^b
{(\alpha^0)^s}_c  \; .
\end{eqnarray}

To consider an $n$-Leibniz deformation of an $n$-Lie algebra one looks
at the Filippov algebra $\mathfrak{G}$ as an $n$-Leibniz one
$\mathfrak{L}$. If the deformed $n$-Leibniz algebra $\mathfrak{L}$ is
required to have a fully antisymmetric $n$-bracket, then we are
actually deforming FAs, and the answer is known: all
semisimple FAs are rigid due to the Whitehead
Lemma for $n$-Lie algebras \cite{FAcoho}, which holds for any $n\geq
2$. Rather than allowing for a general Leibniz bracket we will relax
mildly the full skewsymmetry of the FA $n$-bracket by restricting it
to its first $n-1$ arguments. As mentioned, this is a) natural,
since it keeps the antisymmetry in the arguments of the fundamental
objects (recall that
$\mathscr{X}\cdot Z =[X_1,\dots X_{n-1},Z]$) and b) convenient, since 3-brackets
that are antisymmetric in the first two arguments only have been
used to define the `relaxed three algebras' \cite{Cher-Sa:08,Cher-Do-Sa:08}
that have appeared in the context of the BLG model and which correspond
for $n=3$ to the class of $n$-Leibniz algebras
considered here (see further \cite{Bag-Lam:08} for the `hermitean'
case). In coordinates, this means that the one-cocycles for the
present cohomology problem have the form ${(\alpha^1)^b}_{a_1\dots
a_{n-1}c}$, where only skewsymmetry in the $a_1\dots a_{n-1}$
indices is required. This is what will characterize the possible
deformations of a FA $\mathfrak{G}$ considered as an $n$-Leibniz algebra
of the above type. As stated, this corresponds to having
$\mathscr{X} \in \wedge^{n-1}\mathfrak{L}$ in the resulting
$n$-Leibniz algebra, as it is always the case for FAs.

\subsection{Dualized cocycles}
\label{sec-dual}

This above skewsymmetry restriction allows us to take the dual
$\bar{\alpha}^1$ of $\alpha^1$ by defining
\begin{eqnarray}
\label{forleib11}
{(\bar{\alpha}^1)^b}_{b_1b_2c} &=& \frac{1}{(n-1)!}
{\epsilon^{a_1\dots a_{n-1}}}_{b_1b_2} {(\alpha^1)^b}_{a_1\dots
a_{n-1}c} \ , \nonumber\\
   {(\alpha^1)^b}_{a_1\dots a_{n-1}c} &=&
\frac{1}{2}{\epsilon_{a_1\dots a_{n-1}}}^{b_1b_2}
{(\bar{\alpha}^1)^b}_{b_1b_2c} ; ,
\end{eqnarray}
which will be useful for calculational purposes. Note that the dual
$\bar{\alpha}^1$ of $\alpha^1$ above always has four indices
independently of $n$ and that it is $b_1,b_2$ skewsymmetric,
${(\bar{\alpha}^1)^b}_{b_1b_2c} =
-{(\bar{\alpha}^1)^b}_{b_2b_1c}$.
The invertibility is possible because, for fixed
$b$ and $c$, both $\alpha^1$ and $\bar{\alpha}^1$ have
the same $\left( \begin{array}{c} n+1 \\ n-1 \\
\end{array}\right) = \left( \begin{array}{c} n+1 \\ 2 \\
\end{array} \right) $ degrees of freedom.

  In terms of $\bar{\alpha}^1$,  the cocycle condition
\eqref{forleib9} now reads
\begin{eqnarray}\label{forleib12}
 & & \varepsilon_d {\epsilon_{a_1\dots a_{n-1}e}}^d
{\epsilon_{b_1\dots b_{n-1}}}^{c_1c_2}
{(\bar{\alpha}^1)^e}_{c_1c_2c}
\nonumber\\
& & + \varepsilon_e {\epsilon_{a_1\dots a_{n-1}}}^{c_1c_2}
{\epsilon_{b_1\dots b_{n-1}c}}^e {(\bar{\alpha}^1)^d}_{c_1c_2e}
\nonumber\\
& & -\sum^{n-1}_{r=1} \varepsilon_d {\epsilon_{b_1\dots
b_{r-1}eb_{r+1} \dots b_{n-1}c}}^d {\epsilon_{a_1\dots
a_{n-1}}}^{c_1c_2} {(\bar{\alpha}^1)^e}_{c_1c_2b_r}
\nonumber\\
& & - \varepsilon_d {\epsilon_{b_1\dots b_{n-1}e}}^d
{\epsilon_{a_1\dots a_{n-1}}}^{c_1c_2}
{(\bar{\alpha}^1)^e}_{c_1c_2c}
\nonumber\\
& & -\sum^{n-1}_{r=1} \varepsilon_e {\epsilon_{b_1\dots
b_{r-1}eb_{r+1} \dots b_{n-1}}}^{c_1c_2} {\epsilon_{a_1\dots
a_{n-1}b_r}}^e {(\bar{\alpha}^1)^d}_{c_1c_2c} \nonumber\\
& & - \varepsilon_e {\epsilon_{b_1\dots b_{n-1}}}^{c_1c_2}
{\epsilon_{a_1\dots a_{n-1}c}}^e {(\bar{\alpha}^1)^d}_{c_1c_2e} =0
\ .
\end{eqnarray}
This expression is really
$\delta\alpha^1(\mathscr{X},\mathscr{Y},Z)=0$ for $\mathscr{X}=
(X_{a_1},\dots , X_{a_{n-1}})$, $\mathscr{Y}= (X_{b_1},\dots ,
X_{b_{n-1}})$, so it must be skewsymmetric in $a_1, \dots ,a_{n-1}$
and in $b_1, \dots ,b_{n-1}$. Thus, without losing information, we
can use equally the contraction of (\ref{forleib12}) with
${\epsilon^{a_1 \dots a_{n-1}}}_{a_1'a_2'}{\epsilon^{b_1 \dots
b_{n-1}}}_{b_1'b_2'}$. We obtain, after raising the free index $c$,
the equivalent one-cocycle condition
\begin{eqnarray}\label{forleib13}
& & \varepsilon_d \delta^d_{a_2'}
{(\bar{\alpha}^1)_{a_1'b_1'b_2'}}^c - \varepsilon_d
\delta^d_{a_1'} {(\bar{\alpha}^1)_{a_2'b_1'b_2'}}^c +
\varepsilon_{b_2'} \delta^c_{b_1'}
{(\bar{\alpha}^1)^d}_{a_1'a_2'b_2'} - \varepsilon_{b_1'}
\delta^c_{b_2'}
{(\bar{\alpha}^1)^d}_{a_1'a_2'b_1'}  \nonumber\\
& & - \varepsilon_d \delta^{cd}_{b_1'b_2'}
{(\bar{\alpha}^1)_{ea_1'a_2'}}^e - \varepsilon_d \delta^c_{b_2'}
{(\bar{\alpha}^1)_{b_1'a_1'a_2'}}^d + \varepsilon_d
\delta^c_{b_1'}
{(\bar{\alpha}^1)_{b_2'a_1'a_2'}}^d \nonumber\\
& & + \varepsilon_{b_1'} \delta_{b_1'a_2'}
{{(\bar{\alpha}^1)^d}_{a_1'b_2'}}^c -\varepsilon_{b_1'}
\delta_{b_1'a_1'} {{(\bar{\alpha}^1)^d}_{a_2'b_2'}}^c
-\varepsilon_{b_2'} \delta_{b_2'a_2'}
{{(\bar{\alpha}^1)^d}_{a_1'b_1'}}^c + \varepsilon_{b_2'}
\delta_{b_2'a_1'} {{(\bar{\alpha}^1)^d}_{a_2'b_1'}}^c \nonumber\\
& & - \varepsilon_{a_2'} \delta^c_{a_1'}
{(\bar{\alpha}^1)^d}_{b_1'b_2'a_2'} + \varepsilon_{a_1'}
\delta^c_{a_2'} {(\bar{\alpha}^1)^d}_{b_1'b_2'a_1'} = 0 \;.
\end{eqnarray}

  We may now extract consequences from this condition by taking
different sums of indices. In particular one may, for
instance, contract first $a_1'$ with $b_1'$ after multiplying by
$\varepsilon_{a_1'}$. Then, contracting in  the resulting equation
 (a) $c$ with $d$,
(b) $a_2'$ with $b_2'$ (after multiplying by $\varepsilon{a_2'}$)
and (c) $d$ with $b_2'$ (after multiplying by $\varepsilon_d$), we
obtain the following three expressions
\begin{eqnarray}\label{forleib14}
  &a) & {(\bar{\alpha}^1)_{cab}}^c=0 \; ;\nonumber \\
  &b) & {(\bar{\alpha}^1)_{abc}}^c = {(\bar{\alpha}^1)_{bac}}^c \; ;
  \nonumber\\
 & c)& n\varepsilon_d {{{(\bar{\alpha}^1)_d}^d}_{a_2'}}^c
 +\varepsilon_c {{(\bar{\alpha}^1)^c}_{a_2'd}}^d - \delta^c_{a_2'}
 \varepsilon_e {{(\bar{\alpha}^1)^e}_{ed}}^d = 0\; .
\end{eqnarray}
Note that  equations b) and c) above imply $\varepsilon_c
\varepsilon_e {(\bar{\alpha}^1)_e}^{edc}= \varepsilon_c
\varepsilon_e {(\bar{\alpha}^1)_e}^{ecd}$. Define a new quantity,
$\widetilde{\alpha}^1$,  by
\begin{equation}\label{forleib15}
 {(\widetilde{\alpha}^1)^a}_{bcd} \equiv {(\bar{\alpha}^1)^a}_{bcd} -
 \frac{1}{n} \delta_{cd} {{(\bar{\alpha}^1)^a}_{be}}^e +
 \frac{1}{n} \delta_{bd} {{(\bar{\alpha}^1)^a}_{ce}}^e \ .
\end{equation}
The $\widetilde{\alpha}^1$ above has the following properties: (a) its
traces, $\varepsilon_a {(\widetilde{\alpha}^1)^a}_{acd}$,
${(\widetilde{\alpha}^1)^a}_{bca}$ and ${{(\widetilde{\alpha}^1)^a}_{bc}}^c$
vanish, and (b) $\widetilde{\alpha}^1$ is a one cocycle (it satisfies
\eqref{forleib13}) cohomologous with $\bar{\alpha}^1$. Condition (a)
is true because of (\ref{forleib14}), and (b) follows  because
${(\widetilde{\alpha}^1)^a}_{bcd} - {(\bar{\alpha}^1)^a}_{bcd}\,$, as given
by (\ref{forleib15}), is a one-coboundary. The quickest way to prove
this is to show that any one-cocycle of the form
$(\bar{\beta}^1)_{abcd}  = \delta_{bd}B_{ac} - \delta_{cd}B_{ab}$
with $B_{ab}=B_{ba}$ (in our case
$B_{ab}=\frac{1}{n}{(\bar{\alpha}^1)_{abe}}^e$ which, by the second
equation in \eqref{forleib14}, is $(a,b)$ symmetric) is trivial. To this
aim, we write the corresponding dual one-cocycle $(\beta^1)_{ab_1
\dots b_{n-1}d}$ by (\ref{forleib11}). One finds that in this case
\begin{eqnarray}\label{forleib20}
  (\beta^1)_{ab_1 \dots b_{n-1}d} &\propto & {\epsilon_{b_1 \dots
  b_{n-1}}}^{bc} \bar{\beta}_{abcd} \nonumber\\
  &=& {\epsilon_{b_1 \dots
  b_{n-1}}}^{bc} (\delta_{bd}B_{ac}-\delta_{cd}B_{ab}) \nonumber
  \\
  &=&2 {\epsilon_{b_1 \dots
  b_{n-1}d}}^c B_{ac}\ .
\end{eqnarray}
This expression is a one-cocycle, as can be easily checked, and
moreover one that is totally skewsymmetric in the $n$ indices $b_1,
\dots, b_{n-1},d$, so that it corresponds to the {\it Filippov
algebra} deformations of simple FA's. But since the Whitehead lemma
holds for all $n\geq 2$ simple FAs \cite{FAcoho}, the $\beta^1$ part
of the one-cocycle $\widetilde{\alpha}^1$ is trivial and can be removed
from the deformation.

Now we analyze the consequences of (\ref{forleib13}) for the
traceless $\widetilde{\alpha}^1$ by taking suitable contractions.
First, one may contract $c$ with $b_1'$ in (\ref{forleib13}) with
$\widetilde{\alpha}^1$ instead of $\bar{\alpha}^1$. Then, using that
the contractions of $\widetilde{\alpha}^1$ vanish, one obtains
\begin{equation}\label{forleib16}
  \varepsilon_{b_2'} {(\widetilde{\alpha}^1)^d}_{a_1'a_2'b_2'}
+ \varepsilon_{d} {(\widetilde{\alpha}^1)_{b_2'a_1'a_2'}}^d = 0\; .
\end{equation}
Another possibility is to contract $d$ with $a_2'$ after
multiplying by $\varepsilon_d$, which gives
\begin{equation}\label{forleib17}
  (n-1) {(\widetilde{\alpha}^1)_{a_1'b_1'b_2'}}^c +
{(\widetilde{\alpha}^1)_{b_1'a_1'b_2'}}^c
-{(\widetilde{\alpha}^1)_{b_2'a_1'b_1'}}^c = 0 \ .
\end{equation}
Let us look at this expression closer by taking the cyclic
permutations of (\ref{forleib17}) in in the first three indices.
Lowering the superscript we obtain three equations,
\begin{eqnarray}\label{forleib17a}
   (n-1) \widetilde{\alpha}_{abcd} &=& - \widetilde{\alpha}_{bacd} +
   \widetilde{\alpha}_{cabd}               \nonumber\\
  (n-1) \widetilde{\alpha}_{cabd} &=& - \widetilde{\alpha}_{acbd} +
   \widetilde{\alpha}_{bcad}               \nonumber\\
   (n-1) \widetilde{\alpha}_{bcad} &=& - \widetilde{\alpha}_{cbad} +
   \widetilde{\alpha}_{abcd} \ .
\end{eqnarray}
This yields a homogeneous linear system of three equations and
three unknowns, $\widetilde{\alpha}_{abcd}$, $\widetilde{\alpha}_{bacd}$
and $\widetilde{\alpha}_{cabd}$ (because $\widetilde{\alpha}_{abcd}= -
\widetilde{\alpha}_{acbd}$, see (\ref{forleib11})). By solving this
system one finds
\begin{eqnarray}\label{forleib18}
     (\widetilde{\alpha}^1)_{bacd} & = & - (\widetilde{\alpha}^1)_{abcd}
\nonumber\\
   (3-n) (\widetilde{\alpha}^1)_{abcd} & = & 0 \ .
\end{eqnarray}
We see that for $n\neq 3$ $\widetilde{\alpha}^1$ vanishes. On the
other hand, the first equation of (\ref{forleib18}), together with
(\ref{forleib16}), implies that for $n =3$
$\varepsilon_d(\widetilde{\alpha}^1)_{abcd}$
is completely antisymmetric, hence proportional to $\epsilon_{abcd}$.
It is then straightforward to check that ${(\bar{\alpha}^1)^a}_{bcd}=
\varepsilon_d {\epsilon^a}_{bcd}$ is a solution of eq.
(\ref{forleib13}) for $n=3$.

\subsection{The $n=3$ case}
\label{n=3}

We have shown in Sec.~\ref{sec-dual} above that the most general
one-cocycle in the cohomology for the infinitesimal $n$-Leibniz
deformations (of the particular type specified above) of the $n>2$
simple Filippov algebras is trivial except for $n=3$, in which case
\begin{equation}
\label{forleib19}
   (\bar{\alpha}^1)_{abcd}  =  t \varepsilon_d\epsilon_{abcd}
   \ .
\end{equation}
We now prove that in this $n=3$ case $\bar{\alpha}^1$ above is
non-trivial since it cannot be generated by a zero-cochain. To see
this, we rewrite the triviality condition of eq. \eqref{forleib10}
including the last term in the sum as follows:
\begin{equation}
\label{forleib21}
 {(\alpha^1)^b}_{a_1\dots a_n} =  -\varepsilon_s {(\alpha^0)^b}_s
{\epsilon_{a_1\dots a_n}}^s
 + \sum^{n}_{r=1} \varepsilon_b
{\epsilon_{a_1\dots a_{r-1}sa_{r+1} \dots a_n}}^b
{(\alpha^0)^s}_{a_r} \ .
\end{equation}
The first term is fully skewsymmetric in $a_1\ldots a_n$. The
skewsymmetry of the second one follows from the identity
$\epsilon_{[ a_1 \dots a_n b} {(\alpha^0)^s}_{s]}=0$,
which implies
\begin{equation}\label{forleib23}
    \sum^{n}_{r=1}
{\epsilon_{a_1\dots a_{r-1}sa_{r+1} \dots a_n}}^b
{(\alpha^0)^s}_{a_r} = \epsilon_{a_1\dots a_nb} {(\alpha^0)^s}_s -
\epsilon_{a_1\dots a_ns} {(\alpha^0)^s}_b \; .
\end{equation}
This shows that the second term in \eqref{forleib21} is also
skewsymmetric in the $a_1,\dots ,a_n$ arguments. Thus, any
one-coboundary of the simple FA is necesarily skewsymmetric.
In in our case, however, $\bar{\alpha}^1$ gives,
using \eqref{forleib11},
\begin{equation}
\label{forleib24}
    \alpha^1_{abcd} \propto {\epsilon_{bc}}^{b'c'}
    \varepsilon_d \epsilon_{ab'c'd}= 2(\delta_{ba}\delta_{cd}-
    \delta_{bd}\delta_{ca})\varepsilon_d  \; ,
\end{equation}
which is not skewsymmetric in $b,c,d$ and, therefore, is a
non-trivial one-cocycle. For the Euclidean case, this recovers
the $A_4$ deformation given in \cite{F-O'F:08}.

\subsection{The simple 2-Lie (Lie) algebras case and general results}
\label{Leib-def-of-Lie}

   The proof above clearly applies to the Euclidean
2-Lie algebra $A_3$ {\it i.e.}, to $so(3)$ (as well as to
its pseudoEuclidean version $so(1,2)$), but it does
not extend to the other simple 2-Lie algebras of
the Cartan classification. Nevertheless, it is easy to
prove explicitly that the above result remains true for the
Leibniz algebra deformations of all the $n=2$ simple Filippov
algebras {\it i.e.}, that {\it simple (in fact, semisimple)
Lie algebras remain Leibniz rigid when viewed as Leibniz algebras}.\\

To this end, we first write the one-cocycle condition
\eqref{forleib3} for $n=2$,
\begin{eqnarray}
\label{n2cocycle}
   & &  f_{ae}{}^d (\alpha^1)^e{}_{bc}
  + (\alpha^1)^d{}_{ae} f_{bc}{}^e
- f_{ec}{}^d (\alpha^1)^e{}_{ab} \nonumber\\
& & - f_{be}{}^d (\alpha^1)^e{}_{ac} - (\alpha^1)^d{}_{ec}
f_{ab}{}^e - (\alpha^1)^d{}_{be} f_{ac}{}^e = 0 \ .
\end{eqnarray}
Let us separate the symmetric and antisymmetric parts of the
Leibniz algebra cocycle $\alpha^1$, $(\alpha^1)^e{}_{bc}
 = (\alpha_S^1)^e{}_{bc}
+ (\alpha_A^1)^e{}_{bc}$, $(\alpha_S^1)^e{}_{bc} =
(\alpha_S^1)^e{}_{cb}$, $(\alpha_A^1)^e{}_{bc} =
-(\alpha_A^1)^e{}_{cb}$. We show now that, if
$f_{ab}{}^e$ are the structure constants of a {\it semisimple} Lie
algebra $\mathfrak{g}$, then $(\alpha_S^1)^e{}_{bc}=0$. Indeed, taking the
symmetric part in the indices $a,b$ of eq.~\eqref{n2cocycle}, we
arrive at
\begin{equation}
\label{n2symm}
  f_{ec}{}^d (\alpha^1_S)^e{}_{ab}= 0 \ .
\end{equation}
Contracting this expression with $ f_{e'd}{}^c$ we obtain $k_{e'
e}  (\alpha^1_S)^e{}_{ab}=0$, where $k_{e' e}$ are the coordinates
of the Cartan-Killing metric. Since $\mathfrak{g}$ is semisimple,
$k$ is non-degenerate and necessarily
$(\alpha^1_S)^e{}_{ab}=0$. Therefore, $\alpha^1$ is
antisymmetric and hence a one-cocycle for the
deformation problem of semisimple {\it Lie} algebras.
Since these deformations are all trivial by virtue of the
Whitehead Lemma, it follows that there are no non-trivial
Leibniz deformations of semisimple Lie algebras.\\

Collecting the above $n\geq 2$ results, we have thus proved the following

\begin{theorem}
The $n$-Leibniz algebra deformations of the $(n+1)$-dimensional
simple FA's that preserve the skewsymmetry of the $(n-1)$ first
elements in the $n$-Leibniz bracket (or that of the fundamental
objects) are all trivial for $n > 3$. Further, all $n=2$
semisimple Filippov ({\it i.e.}, Lie) algebras are rigid
as Leibniz algebras.
\end{theorem}

  One may ask what makes the simple FA $n=3$  deformation case special. In
the present context this is simply answered by noticing that, for the class of
$n$-Leibniz deformations we are considering, a one-cocycle with a dual given
by the four index Levi-Civita fully antisymmetric symbol may exist
only for a four dimensional simple Filippov algebra. Since all
the simple real Filippov algebras are mixed signature versions of
the Euclidean FA $A_{n+1}$ and have dimension $n+1$, this gives
$n=3$.

\section{A class of $n$-Leibniz central extensions of simple FAs}

\label{Leib-central-Lie}

Let us now move to the case of the $n$-Leibniz central extensions
of the simple FAs when the resulting Leibniz algebra is of the type
considered in Sec.~\ref{n-L-defSec}, that is, one with an $n$-bracket
required to be skewsymmetric in its first $n-1$ entries.

As stated in Sec.~\ref{n-L-central}, all formulae relevant for the central extension
cohomology may easily be derived from those in Secs.~\ref{Leib-def-FAn>2},
\ref{sec-dual} by taking $\mathbb{R}$-valued cochains {\it i.e.}, by removing
the first upper index and by eliminating from the expression of the action of the
coboundary operator $\delta$ the terms containing the action on the
$\mathfrak{L}$-valued cochains. In this way, the dualized coordinate
expression for the one-cocycle and one-coboundary conditions become,
respectively,
\begin{eqnarray}
\label{Pext1}
 & & \varepsilon_{b_2'} \delta^c_{b_1'} (\bar{\alpha}^1)_{a_1'a_2'b_2'}
- \varepsilon_{b_1'} \delta^c_{b_2'}
(\bar{\alpha}^1)_{a_1'a_2'b_1'}
\nonumber\\
& &+ \varepsilon_{b_1'} \delta_{b_1'a_2'}
{(\bar{\alpha}^1)_{a_1'b_2'}}^c - \varepsilon_{b_1'}
\delta_{b_1'a_1'} {(\bar{\alpha}^1)_{a_2'b_2'}}^c
\nonumber\\
& & - \varepsilon_{b_2'} \delta_{b_2'a_2'}
{(\bar{\alpha}^1)_{a_1'b_1'}}^c + \varepsilon_{b_2'}
\delta_{b_2'a_1'} {(\bar{\alpha}^1)_{a_2'b_1'}}^c
\nonumber\\
& & - \varepsilon_{a_2'} \delta^c_{a_1'}
(\bar{\alpha}^1)_{b_1'b_2'a_2'} + \varepsilon_{a_1'}
\delta^c_{a_2'} (\bar{\alpha}^1)_{b_1'b_2'a_1'} = 0
\end{eqnarray}
and
\begin{equation}
\label{Pext2} {(\delta \bar{\alpha}^1)_{b_1b_2}}^c = -
\varepsilon_{b_2} \delta_{b_1}^c (\alpha^0)_{b_2} +
\varepsilon_{b_1} \delta_{b_2}^c (\alpha^0)_{b_1} \quad ,
\end{equation}
where, in a way analogous to eq. (3.34), $\bar{\alpha}^1$ is given
by
\begin{equation}
\label{Pext3}
  \bar{\alpha}^1_{b_1b_2c}  = \frac{1}{(n-1)!} \epsilon^{b_1b_2a_1 \dots
a_{n-1}} {\alpha}^1_{b_1\dots a_{n-1}c} \ .
\end{equation}
Given an $\bar{\alpha}^1$ that satisfies the one-cocycle condition
\eqref{Pext1}, we now consider the equivalent cocycle
$\widetilde{\alpha}^1$ given by
\begin{equation}
\label{Pext4}
     {(\widetilde{\alpha}^1)_{b_1b_2}}^c = {(\bar{\alpha}^1)_{b_1b_2}}^c
     - \frac{1}{n} \delta^c_{b_1} {(\bar{\alpha}^1)_{eb_2}}^e +\frac{1}{n}
\delta^c_{b_2} {(\bar{\alpha}^1)_{eb_1}}^e
  \ ,
\end{equation}
where the last two terms define a one-coboundary generated by
$\alpha^0=\frac{1}{2}\varepsilon_{b_2} {(\bar{\alpha}^1)_{eb_2}}^e$
(see \eqref{Pext2}). This new cocycle $\widetilde{\alpha}^1$ also obeys
eq.~\eqref{Pext1}, and has the property
that the trace ${(\widetilde{\alpha}^1)_{b_1c}}^c$ vanishes. In this
way, if one contracts $b_1'$ with $c$ in eq. \eqref{Pext1} for
$\widetilde{\alpha}^1$ and takes into account the vanishing of the
trace, one arrives at
\begin{equation}
\label{Pext5}
        n {(\widetilde{\alpha}^1)_{b_1b_2}}^c = 0\ ,
\end{equation}
which shows that $\widetilde{\alpha}^1=0$ and hence that all
one-cocycles are trivial for any $n$. We have thus proved the
following

\begin{theorem}
The $n$-Leibniz algebra central extensions  of simple FA's that
preserve the skewsymmetry of the $(n-1)$ first entries of the
$n$-bracket (or of the fundamental objects) are all trivial for any
$n > 2$.
\end{theorem}
\noindent
The proof above also applies to the 2-Lie simple algebras $A_3\; (so(3))$ and
$so(1,2)$; for general simple Lie algebras, see \cite{Lod-Pir:96} and
\cite{Hu-Pei-Liu:06} (Prop. 3.2 and Cor. 3.7).

\section{Conclusions}
\label{final-remarks}

We have shown that the simple Filippov algebras, when viewed as
$n$-Leibniz algebras of the class that have $n$-brackets
antisymmetric in the first $n-1$ entries (and thus skewsymmetric
fundamental objects), are rigid but for $n=3$ (Theorem 1). Further, simple
$n$-Lie algebras do not have $n$-Leibniz central extensions
within the same class (Theorem 2).

Obviously, there is the question of whether a further relaxing of
the full skewsymmetry condition, {\it i.e.}, whether allowing for more
general $n$-Leibniz brackets, results in more deformations or
non-trivial central extensions. In the case of deformations, we expect that
the situation will change, because the natural cochains dual to
those with less than $n-1$ antisymmetric entries have more than four
indices, and Levi-Civita symbols with five or more indices may lead
to additional non-trivial $n$-Leibniz deformations of the simple
Filippov algebras.

\subsection*{Acknowledgments}

This work has been partially supported by research grants from the
Spanish Ministry of Science and Innovation (FIS2008-01980,
FIS2009-09002) and the Junta de Castilla y Le\'on (VA-013-C05).

\end{document}